\documentclass[twocolumn,aps,prd,showpacs,amsmath,amssymb]{revtex4}
\usepackage[]{latexsym}
\usepackage[english]{babel}
\usepackage{graphicx}

\begin{document}

\title{Updated constraints on $f(\mathcal{R})$ gravity from cosmography}

\author{Alejandro Aviles}
\email{aviles@ciencias.unam.mx}
\affiliation{Depto. de F\'isica, Instituto Nacional de Investigaciones Nucleares, M\'exico, Mexico}
\affiliation{Instituto de Ciencias Nucleares, UNAM, M\'exico, Mexico.}

\author{Alessandro Bravetti}
\email{bravetti@icranet.org}
\affiliation{Dip.  di Fisica and ICRA, "Sapienza" Universit\`a di Roma,
             Piazzale Aldo Moro 5, I-00185, Roma, Italy}
\affiliation{Instituto de Ciencias Nucleares, UNAM, M\'exico, Mexico.}

\author{Salvatore Capozziello}
\email{capozzie@na.infn.it}
\affiliation{Dip. di Scienze Fisiche, Universit\`a di Napoli "Federico II",
             Via Cinthia, I-80126, Napoli, Italy}
\affiliation{INFN Sez. di Napoli, Compl. Univ. Monte S. Angelo Ed. N
Via Cinthia, I- 80126 Napoli, Italy.}

\author{Orlando Luongo}
\email{orlando.luongo@na.infn.it}
\affiliation{Instituto de Ciencias Nucleares, UNAM, M\'exico, Mexico;}
\affiliation{Dip. di Scienze Fisiche, Universit\`a di Napoli "Federico II",
             Via Cinthia, I-80126, Napoli, Italy}
\affiliation{INFN Sez. di Napoli, Compl. Univ. Monte S. Angelo Ed. N
Via Cinthia, I- 80126, Napoli, Italy.}

\begin{abstract}
We address the issue of constraining the class of $f(\mathcal{R})$ able to reproduce the observed cosmological acceleration,  by using the so called cosmography of the universe.  We consider a model independent procedure to build up a $f(z)$-series in terms of the measurable cosmographic coefficients; we therefore derive cosmological late time bounds on $f(z)$ and its derivatives up to the fourth order, by fitting the luminosity distance \emph{directly} in terms of such coefficients. We perform a Monte Carlo analysis, by using three different statistical sets of cosmographic coefficients, in which the only assumptions are the validity of the cosmological principle and that the class of $f(\mathcal{R})$ reduces to $\Lambda$CDM when $z\ll1$. We use the updated union 2.1 for supernovae Ia, the constrain on the $H_0$ value imposed by the measurements of the Hubble space telescope and the Hubble dataset, with measures of $H$ at different $z$. We find a statistical good agreement of the $f(\mathcal{R})$ class under exam, with the cosmological data; we thus propose a candidate of $f(\mathcal{R})$, which is able to pass our cosmological test, reproducing the late time acceleration in agreement with observations.
\end{abstract}

\pacs{04.50.+h, 04.20.Ex, 04.20.Cv, 98.80.Jr}
\maketitle

\section{Introduction}

The recent observational evidence of the late time acceleration of the universe \cite{SNeIa,zump}
opened new challenges in the framework of theoretical cosmology. To explain the origin of such a cosmic speed up, cosmologists usually assume the existence of an exotic fluid called dark energy (DE) \cite{copeland}.
Even though its physical nature is still unclear, several attempts have been
made in order to resolve the problem of its existence \cite{dopocopeland}.
In general relativity (GR),
the simplest possibility is the introduction of a vacuum energy cosmological constant, $\Lambda$ \cite{Lambda,againstlambda}.
The resulting model is usually referred to as $\Lambda$CDM \cite{au}.
However, alternative approaches have followed each other, without being conclusive
\cite{more1,teorie,mine,mine2}.
To this regard, another appealing possibility is to consider GR
as a limiting theory of a more general paradigm \cite{rev1}; so that, in the last decades, particular attention has been devoted to solve the problem of the accelerated universe in the framework of extended theories of gravity \cite{rev2}.
Generally, extending GR means to review the DE effects as due to possible corrections of the Einstein-Hilbert action \cite{rev3}.

In this paper, we focus our attention to the case of the so called $f(\mathcal{R})$ theories,
in which the Ricci scalar $\mathcal{R}$ in the Einstein-Hilbert action
is replaced by a more general analytic function, namely $f(\mathcal{R})$.
The corresponding action reads
${\cal{A}} = \int{d^4x \sqrt{-g} \left [ f(\mathcal{R}) +
{\cal{L}}_{m}\right]}$ \cite{rev4},
where ${\cal{L}}_{m}$ is the standard matter term.
By varying the action with respect the metric $g_{\mu\nu}$, we obtain the field equations \cite{rev5}
\begin{equation}\label{filedeqs}
\mathcal{R}_{\mu \nu}f^{'}(\mathcal{R})-\frac{1}{2}f(\mathcal{R})g_{\mu
\nu}-(\nabla_{\mu}\nabla_{\nu}-g_{\mu
\nu}\nabla_{\alpha}\nabla^{\alpha})f^{'}(\mathcal{R}) =8\pi T_{\mu \nu}\,,
\end{equation}
in the case of the metric approach where the
connection is the Christoffel one. Here $T_{\mu\nu}$ is the standard energy momentum tensor and $G=c=1$.

The problem of determining the nature of DE is therefore shifted to understand which $f(\mathcal{R})$
is the correct candidate to explain the dynamics of the universe. The correct class of $f(\mathcal{R})$ should be compatible with modern observations \cite{altro}. Therefore, we propose to limit our attention only to the class of $f(\mathcal{R})$ reducing to $\Lambda$CDM at the low redshift regime  \cite{altrissimo1,altrissimo2,altrissimo3,starobinskievisser,mioprdnuovo}.

In order to check the viability of $f(\mathcal{R})$ models, it has been proposed in \cite{capozziello}
to study the so called \emph{cosmography} of $f(\mathcal{R})$.  Cosmography represents a part of cosmology which does not postulate any cosmological model \emph{a priori}. Thus, it can be thought as a model independent way to fix constraints on the universe dynamics at late times through the use of a set of parameters, namely the cosmographic set (CS). The aim of this work is to relate the $f(\mathcal{R})$ Taylor expansion around $z\sim0$ to the luminosity distance, determining the Taylor coefficients as functions of the CS. Afterwards, we fit the coefficients by \emph{directly} using the luminosity distance; this allows us to overcome the problem of the error propagation, since the $f(\mathcal{R})$ coefficients are measured directly from data. In particular, once obtained the expression of the luminosity distance in terms of $f(z)$ and its derivatives, we perform a Monte Carlo fitting procedure. We obtain at our time stringent numerical intervals for $f(z)$ and its derivatives up to the fourth order and then the corresponding constraints on $f(\mathcal{R})$ and its derivatives. The set of cosmographic $f(z)$ parameters is measured directly from supernovae Ia (SNeIa) data, $\mathcal{H}(z)$ observations, under the bound provided by the Hubble space telescope for $H_0$. Finally, we get a viable candidate of $f(\mathcal{R})$, reconstructing it from the cosmographic test. Such a candidate seems to pass the cosmographic and cosmological tests, extending the $\Lambda$CDM paradigm as a limiting case of a more general theory.

The paper is structured as follows: in Sec. II we develop the main features of cosmography and we define the so called $f(z)$ cosmographic set, which is the set of $f(z)$ and its derivatives to be fitted with the data. In Sec. III we perform a Monte Carlo analysis, based on three statistical models, while in Sec. IV we propose a viable candidate of $f(\mathcal{R})$, compatible with the bounds inferred from our tests. Finally, in Sec. V we develop conclusions and perspectives of our work.

\section{Cosmographic $f(z)$ parameters}

In this section, we relate the $f(\mathcal{R})$ coefficients (evaluated in terms of the redshift $z$) to the cosmographic set (CS). Afterwards, we use these relations to write the luminosity distance in terms of $f(z)$ and its derivatives at $z=0$.
To this end, let us review briefly the theoretical features of cosmography.
Cosmography, or alternatively cosmo-kinetics, is a tool to investigate the dynamics of the universe,
regardless of the particular cosmological model.
Cosmography indeed simply postulates the validity of the cosmological principle. Thus, it follows the use of the Friedmann-Robertson-Walker (FRW) metric, i.e.
\begin{equation}\label{frw}
ds^2=dt^2-a(t)^2\left(dr^2+r^2d\Omega^2\right)\,,
\end{equation}
where we assume hereafter a spatially flat universe ($k=0$) and we use the notation $d\Omega^2\equiv d\theta^2+\sin^2\theta d\phi^2$.

The paradigm of cosmography is to expand the scale factor $a(t)$ in a Taylor series around the present time $t_0$ \cite{Weinberg2008}.
We give here the expressions for the first $6$ coefficients in the expansion,
\begin{eqnarray}\label{pinza}
\mathcal{H} \equiv \frac{1}{a} \frac{da}{dt}\,,\quad&\,& \quad q \equiv -\frac{1}{a\mathcal{H}^2} \frac{d^2a}{dt^2}\,,\nonumber\\
j  \equiv \frac{1}{a\mathcal{H}^3} \frac{d^3a}{dt^3}\,,\quad&\,& \quad s \equiv \frac{1}{a\mathcal{H}^4} \frac{d^4a}{dt^4}\,,\\
l  \equiv \frac{1}{a\mathcal{H}^5} \frac{d^5a}{dt^5}\,,\quad&\,& \quad m \equiv
\frac{1}{a\mathcal{H}^6}\frac{d^6a}{dt^6}\,. \nonumber
\end{eqnarray}
The coefficients in eq. (\ref{pinza}) are, by construction,  model independent quantities,
which are called the cosmographic set (CS). They are known in the literature as the Hubble rate ($\mathcal{H}$), the acceleration parameter ($q$), the jerk parameter ($j$), the snap parameter ($s$), the lerk parameter ($l$) \cite{starobinskievisser} and the $m$ parameter introduced in \cite{mioprdnuovo}. The set of such parameters is known in the literature as the CS.

\subsection{Degeneracy and cosmography}

The definitions given above lead to the most relevant property of cosmography, that is,
 overcoming the so called degeneracy problem among different cosmological models.
In fact, no cosmological model is assumed \emph{a priori} in the expression of the luminosity distance.
Furthermore, another significative aspect of cosmography is to relate the series expansion of the luminosity distance to the CS. To this regard, it was pointed out \cite{where,luongo1,Weinberg2008} that direct measurements of such quantities are permitted, overcoming the problem of the statistical error propagations. Hence, it is possible to compare theoretical predictions with the observed values, without passing through a cyclic scheme which postulates a priori the form of $\mathcal{H}$ and $f(R)$ \cite{altro}.

One of the most important observational quantities to be expanded in series is the luminosity distance $d_L$.
Considering the scale factor definition in terms of $z$,  i. e. $a\equiv(1+z)^{-1}$, the luminosity distance reads
\begin{eqnarray}\label{serie2}
d_L  =   \sqrt{\frac{\mathcal{L}}{4\pi \mathcal{F}}}=\frac{r_0}{a(t)}\,, \nonumber \\
\end{eqnarray}
where we defined $\mathcal{L}$ and $\mathcal{F}$ as the luminosity and the flux respectively, while
\begin{equation}
r_0 = \int_{t}^{t_0}{\frac{d\xi}{a(\xi)}}\,,
\end{equation}
whose physical meaning is related to the distance $r$ that a photon travels from a light source at $r=r_0$ to our position at $r=0$. Equation ($\ref{serie2}$) can be expanded in powers of $z$ around $z=0$;
the expansion up to the sixth order in $z$ is given in the Appendix A, both in terms of the CS and in terms of the derivatives of $f(z)$. Now we want to write $f(\mathcal R)=f(\mathcal R(z))=f(z)$ and use the definitions in ($\ref{pinza}$) to express $f(z)$ in terms of the CS, i.e. $f(z)=f(\mathcal{H}(z),q(z),j(z),s(z),l(z),m(z))$.

\noindent To do so, let us start from the definition of $\mathcal{R}$ in terms of $t$ and $\mathcal{H}$, i.e.
\begin{equation}\label{eq: constr}
\mathcal{R} = -6 \left ( \dot{\mathcal{H}} + 2 \mathcal{H}^2\right )\,.
\end{equation}
Using the redshift definition in terms of the cosmic time
\begin{equation}\label{zdit}
\frac{d\log(1+z)}{dt}=-\mathcal{H}(z)\,,
\end{equation}
we rewrite $\mathcal R$ in terms of $z$, obtaining
\begin{equation}\label{eq: constr}
\mathcal{R} = 6 \left[ (1+z)\mathcal{H}\,\mathcal{H}_{z} - 2 \mathcal{H}^2\right]\,.
\end{equation}
Hence, we can calculate $\mathcal{R}$ and its derivatives in terms of $z$ and evaluate them in $z=0$.
The result, up to
the fifth derivative, is

\begin{equation}\label{Rinz0}
\begin{split}
\frac{\mathcal{R}_0}{6} =\, & \mathcal{H}_0\left[\mathcal{H}_{z0}-2\mathcal{H}_0\right]\,,\\
\frac{\mathcal{R}_{z0}}{6} =\,& \mathcal{H}_{z0}^2+\mathcal{H}_0(-3\mathcal{H}_{z0}+\mathcal{H}_{2z0})\,,\\
\frac{\mathcal{R}_{2z0}}{6} =\,& -2\mathcal{H}_{z0}^2+3\mathcal{H}_{z0}\mathcal{H}_{2z0}+\mathcal{H}_0(-2\mathcal{H}_{2z0}+\mathcal{H}_{3z0})\,,\\
\frac{\mathcal{R}_{3z0}}{6}=\,&3\mathcal{H}_{2z0}^2+\mathcal{H}_{z0}\left(-3\mathcal{H}_{2z0}+4\mathcal{H}_{3z0}\right)
+\mathcal{H}_0(-\mathcal{H}_{3z0}\\
&+\mathcal{H}_{4z0})\,,\\
\frac{\mathcal{R}_{4z0}}{6} =\,&10\mathcal{H}_{2z0}\mathcal{H}_{3z0}+5\mathcal{H}_{z0}\mathcal{H}_{4z0}+\mathcal{H}_0\mathcal{H}_{5z0}\,,\\
\frac{\mathcal{R}_{5z0}}{6} =\,&10\mathcal{H}_{3z0}\big(\mathcal{H}_{2z0}+\mathcal{H}_{3z0}\big)+15\mathcal{H}_{2z0}\mathcal{H}_{4z0}\\
&+\mathcal{H}_{z0}\big(5\mathcal{H}_{4z0}+6\mathcal{H}_{5z0}\big)+\mathcal{H}_0\big(\mathcal{H}_{5z0}+\mathcal{H}_{6z0}\big)\,,
\end{split}
\end{equation}
where, hereafter,  we adopt the convention $\frac{d^nX}{dz^n}\Big|_{0}\equiv X_{nz0}$, for $X$ a generic function of $z$.

Therefore, in order to evaluate $\mathcal{R}=\mathcal{R}(\mathcal{H}_0,q_0,j_0,s_0,l_0,m_0)$,
we need to express  $\mathcal{H}$ and its derivatives in terms of the CS.
To this regard, after some cumbersome algebra, we infer from Eqs. ($\ref{pinza}$)

\begin{eqnarray}\label{eq:CSoftime}
q&=&-\frac{\dot{\mathcal{H}}}{\mathcal{H}^2} -1\,, \nonumber\\
j&=&\frac{\ddot{\mathcal{H}}}{\mathcal{H}^3}-3q-2\,, \nonumber\\
s&=&\frac{\mathcal{H}^{(3)}}{\mathcal{H}^4}+4j+3q\left(q+4\right)+6\,,\\
l&=&\frac{\mathcal{H}^{(4)}}{\mathcal{H}^5}-24 - 60 q - 30 q^2 - 10 j \left(q+2\right) + 5 s\,, \nonumber\\
m&=&\frac{\mathcal{H}^{(5)}}{\mathcal{H}^6} + 10 j^2 + 120 j \left(q+1\right)
+\nonumber\\ &&3 \left[2 l + 5 \left(24 q
 + 18 q^2 + 2 q^3 - 2 s - q s+8\right)\right]\,, \nonumber
\end{eqnarray}

\noindent and then the corresponding derivatives of $\mathcal{H}$ in terms of the cosmic time read

\begin{equation}\label{Hpunto}
\begin{split}
\frac{d\mathcal{H}}{dt} =& -\mathcal{H}^2 (1 + q)\,,\\
\frac{d^{2}\mathcal{H}}{dt^2} =& \mathcal{H}^3 (j + 3q + 2)\,,\\
\frac{d^{3}\mathcal{H}}{dt^3} =& \mathcal{H}^4 \left [ s - 4j - 3q (q + 4) - 6 \right]\,,\\
\frac{d^4\mathcal{H}}{dt^4} =& \mathcal{H}^5 \left [ l - 5s + 10 (q + 2) j + 30 (q + 2) q + 24\right ]\,,\\
\frac{d^5\mathcal{H}}{dt^5} =& \mathcal{H}^6
\Big\{m-4l+12s+7sq-24j-32jq-4j^2\\
&-24q-36q^2-6q^3-6(j+3q+2)^2\\
&+8(1+q)(s-4j-3q(q+4)-6)\\
&-2\big[l-5s+(10j+30q)(q+2)+24\big]\Big\}\,.
\end{split}
\end{equation}

Thus, using Eq. (\ref{zdit}), we can rewrite Eqs. (\ref{Hpunto}) in terms of the CS only, obtaining

\begin{equation}\label{Hinz0}
\begin{split}
\mathcal{H}_{z0}/\mathcal{H}_0=\,& 1+q_0\,,\\
\mathcal{H}_{2z0}/\mathcal{H}_0=\,& j_0-q_0^2\,,\\
\mathcal{H}_{3z0}/\mathcal{H}_0=\,&-3j_0-4j_0q_0+q_0^2+3q_0^3-s_0\,,\\
\mathcal{H}_{4z0}/\mathcal{H}_0=\,&12j_0-4j_0^2+l_0+32j_0q_0-12q_0^2+25j_0q_0^2\\
&-24q_0^3-15q_0^4+8s_0+7q_0s_0\,,\\
\mathcal{H}_{5z0}/\mathcal{H}_0=\,&32j_0q_0-15l_0-11l_0q_0+60q_0^2+180q_0^3\\
&+225q_0^4+105q_0^5+10j_0^2(6+7q_0)-m_0\\
&-j_0(60+272q_0+375q_0^2+210q_0^3-15s_0)\\
&-60s_0-98q_0s_0-60q_0^2s_0-7q_0s_0\,.
\end{split}
\end{equation}

Then, using equations (\ref{Rinz0}) and (\ref{Hinz0}), we are able to evaluate the expressions of $\mathcal{R}$ and its derivatives as functions of the CS only.

\subsection{The use of the modified Friedmann equations}

In this subsection, we want to show the procedure to fix constraints on $f(\mathcal{R})$ and its derivatives. We therefore use Eqs. (\ref{Rinz0}) and (\ref{Hinz0}) and we consider the modified Friedmann equations, derived by assuming the FRW metric and Eq. ($\ref{filedeqs}$).

In the case of the standard matter term, ($\rho_m\propto a^{-3}$ and $P_m=0$), one gets the modified Friedmann equations

\begin{equation}
\mathcal{H}^2 = \frac{1}{3} \left [ \rho_{curv} + \frac{\rho_m}{f'(\mathcal{R})} \right
]\,, \label{eq1}
\end{equation}
and
\begin{equation}\label{eq2}
2 \dot{\mathcal{H}} + 3\mathcal{H}^2= - P_{curv}\,.
\end{equation}

\noindent Equations $(\ref{eq1})$ and (\ref{eq2}) determine the definition of the DE fluid in terms of the curvature as
\begin{equation}\label{eq:rhocurv}
\rho_{curv} = \frac{1}{f'(\mathcal{R})} \left \{ \frac{1}{2} \bigg[ f(\mathcal{R})  - \mathcal{R}
f'(\mathcal{R}) \bigg] - 3 \mathcal{H} \dot{\mathcal{R}} f''(\mathcal{R}) \right \} \,.
\end{equation}
The corresponding barotropic pressure reads
\begin{equation}
P_{curv} = \omega_{curv} \rho_{curv} \label{eq: pcurv}\,,
\end{equation}
with the definition of the effective curvature barotropic factor given by
\begin{equation}
\omega_{curv} = -1 + \frac{\ddot{\mathcal{R}} f''(\mathcal{R}) + \dot{\mathcal{R}} \left [ \dot{\mathcal{R}}
f'''(\mathcal{R}) - \mathcal{H} f''(\mathcal{R}) \right ]} {\left [ f(\mathcal{R}) - \mathcal{R} f'(\mathcal{R}) \right ]/2 - 3
\mathcal{H} \dot{\mathcal{R}} f''(\mathcal{R})}\,. \label{eq: wcurv}
\end{equation}

Assuming the functional dependence $\mathcal{R}=\mathcal{R}(z)$, we rewrite each term of Eq. (\ref{eq: wcurv}) in terms of $z$. We get

\begin{equation}
\begin{split}\label{zuzzu}
f'(\mathcal{R}) =\, & \mathcal{R}_z^{-1}f_z\,,\\
f''(\mathcal{R})=\, & (f_{2z}\mathcal{R}_z - f_z\mathcal{R}_{2z})\mathcal{R}_z^{-3}\,,\\
f'''(\mathcal{R})=\, &\frac{f_{3z}}{\mathcal{R}_z^3} - \frac{f_z\, \mathcal{R}_{3z}+3f_{2z}\,
\mathcal{R}_{2z}}{\mathcal{R}_z^4}+\frac{3f_z\, \mathcal{R}_{2z}^2}{\mathcal{R}_z^5}\,,
\end{split}
\end{equation}
and, using equation (\ref{zdit}),
\begin{equation}\label{rdotrddot}
\begin{split}
\dot{\mathcal{R}}=&-(1+z)\mathcal{H}\mathcal{R}_z\,,\\
\ddot{\mathcal{R}}=&(1+z)\mathcal{H}\big[\mathcal{H}\mathcal{R}_z+(1+z)(\mathcal{H}_z\mathcal{R}_z+\mathcal{H}\mathcal{R}_{2z})\big]\,.
\end{split}
\end{equation}

Furthermore, following \cite{capozziello}, we know that any  $f(\mathcal{R})$ theory requires

\begin{equation}\label{constraint1}
f''(\mathcal{R}_0) = 0\,,
\end{equation}

\noindent in order to be compatible with Solar System tests
and

\begin{equation}\label{constraint2}
f'(\mathcal{R}_0) = 1\,,
\end{equation}

\noindent to predict the correct value for the gravitational constant $G$.

Therefore, combining equation ($\ref{zuzzu}$) with ($\ref{Rinz0}$) and ($\ref{Hinz0}$), we have

\begin{equation}\label{f0fz0fzz0}
\begin{split}
\frac{f_0}{6\mathcal{H}_0^2}=\, &-\Omega_m+q_0\,,\\
\frac{f_{z0}}{6\mathcal{H}_0^2}=\, &\frac{\mathcal{R}_{z0}}{6\mathcal{H}_0^2}=\,-2-q_0+j_0\,,\\
\frac{f_{2z0}}{6\mathcal{H}_0^2}=\,&\frac{\mathcal{R}_{2z0}}{6\mathcal{H}_0^2}=-2-4q_0-(2+q_0)j_0-s_0\,,
\end{split}
\end{equation}

\noindent where we used the condition that $\rho_{curv0} = \frac{f(\mathcal{R}_0) - \mathcal{R}_0}{2}$
and that $f_0  =6 \mathcal{H}_0^2 (1 - \Omega_m) + \mathcal{R}_0$.

Now, using equations (\ref{zuzzu}) and (\ref{rdotrddot}) in (\ref{eq: wcurv}), we can write $\omega_{curv}$
as a function of $z$ only. Then we expand this expression as a Taylor series around $z=0$, obtaining

\begin{equation}\label{wexpanded}
\omega_{curv}=\sum_{j=0}^{\infty}\frac{1}{j!}\frac{d^j\omega_{curv} }{dz^j}\Big|_{z=0}z^j\,.
\end{equation}

The first term in this expansion, which we call $\omega_0$, depends only on $f$ and its derivatives up to the third order (evaluated at $z=0$), while the second term $\omega_1$ depends on $f$ and its derivatives up to the fourth order, and so forth for the higher terms.

Keeping in mind that the class of $f(\mathcal{R})$ should reduce to $\Lambda$CDM at low redshift regime,
we  compare our results with  $\Lambda$CDM;
thus, by fixing in equation ($\ref{wexpanded}$) the $\Lambda$CDM bounds

\begin{eqnarray}\label{w0LCDM}
\Omega_m&=&\frac{2}{3}(1+q_0)\,,\nonumber\\
\omega_0^{\Lambda CDM}&=&-\frac{1}{3}(1-2q_0)\,,\nonumber\\
\omega_1^{\Lambda CDM}&=&0\,,\nonumber\\
\end{eqnarray}

we get $f_0$, $f_{z0}$, $f_{2z0}$, $f_{3z0}$ and $f_{4z0}$ in terms of the CS only

\begin{equation}\label{f0fz0fzz0dopo}
\begin{split}
\frac{f_0}{2\mathcal{H}_0^2}=\,&-2+q_0\,,\\
\frac{f_{z0}}{6\mathcal{H}_0^2} =\,&-2-q_0+j_0\,,\\
\frac{f_{2z0}}{6\mathcal{H}_0^2}=\,&-2-4q_0-(2+q_0)j_0-s_0\,,\\
\frac{f_{3z0}}{2\mathcal{H}_0^2} =\,& -4 - 3 j_0^2 + 3 l_0 + j_0 \big[2 + q_0 (13 + 5 q_0)\big]\\
& + 15 s_0 +
 q_0 \big[2 + 2 q_0 (5 + 2 q_0) + 9 s_0\big]\,,\\
\frac{f_{4z0}}{2\mathcal{H}_0^2} =\,&
8 + 30 j_0^2 (1 + q_0) - 6 l_0 (5 + 3 q_0) - 3 m_0- 66 s_0 \\
&- j_0 \big[22 + q_0 \big(46 + q_0 (38 + 29 q_0)\big) - 15 s_0\big] \\
&- q_0 \big[18 + 84 s_0 + q_0 \big(4 + 2 q_0 (-9 + 2 q_0) \\
&+ 33 s_0\big)\big]\,.
\end{split}
\end{equation}

\noindent We refer to Eqs. ($\ref{f0fz0fzz0dopo}$) as the definition of the $f(z)$-cosmographic set (fCS).
Now our intent is to constrain the values of  $f_0$, $f_{z0}$, $f_{2z0}$, $f_{3z0}$ and $f_{4z0}$.
To do so, we write the luminosity distance $d_L$ in terms of  the fCS by using Eqs. (\ref{f0fz0fzz0dopo}).
This is performed in two steps; first we invert the algebraic system  (\ref{f0fz0fzz0dopo}) to find the CS in terms of the fCS. Then we insert these expressions in equation (\ref{sestoordine}).
The result is $d_L$ as a power series of $z$, whose coefficients are now in terms of the fCS, instead of the CS.
The explicit expression of $d_L$ in terms of the fCS is given in Eq.  (\ref{sestoordinebis}).

In addition, in order to measure the fCS  using $d_L$ and  the cosmological data, we need to define viable priors, compatible with the observed universe. To infer our priors we assume that the class of $f(\mathcal{R})$ reduces to $\Lambda$CDM at late times, as already stressed above. We write such priors in Tab. I.

\newlength{\mywidth}
\setlength{\mywidth}{0.4\textwidth}
\begin{table}
\begin{center}
\begin{tabular}{c}
\begin{tabular*}{\mywidth}{c}
\hline
Flat priors\\ \hline \hline
\begin{tabular}{rcl}
$0.5\quad <$ &$h$ & $< \quad 0.9$\\
$0.001 \quad <$&$\Omega_{\rm b}h^2$ & $< \quad 0.09$ \\
$\,-5 \quad <$&$ 10^{-4} f_0 $&$< \quad  5$ \\
$\,-10 \quad <$&$ 10^{-5} f_{z0} $&$< \quad 10$ \\
$\, -15 \quad <$&$ 10^{-5} f_{2z0} $&$< \quad 15$ \\
$\, -20\quad  <$&$ 10^{-5}  f_{3z0} $&$< \quad 20 $ \\
$\, -50\quad <$&$ 10^{-6}  f_{4z0} $&$< \quad 50$ \\
\end{tabular}\\\hline\\  \hline
\hline
Additional constraints\\ \hline \hline
\begin{tabular}{rl}
$\Omega_{k}$&$ =0$\\
$\omega_m$&$ =0.274$\\
$w_{j}$&$ =0$\\
\end{tabular}\\ \hline
\end{tabular*}
\end{tabular}
\end{center}
\caption{Priors imposed on the parameters in the Monte Carlo analysis.}\label{tab:priors}
\end{table}

We can now perform a best fit for the values of the fCS and obtain constraints on the values of $f(z)$ and its derivatives at present time. This will be the content of the following section.

\section{Monte Carlo analysis and constraints on fCS}

In this section we evaluate the cosmological constraints on the fCS
by fitting the luminosity distance (\ref{sestoordinebis}) with the cosmological data.
We analyze three statistical models with different maximum order of parameters; this procedure, widely adopted in the literature, corresponds to assume a hierarchy among parameters. The sets that we are going to analyze are summarized as

\begin{eqnarray}
 \text{A} &=& \{\mathcal{H}_0, f_0, f_{z0}, f_{2z0}\}\,, \\
 \text{B} &=&\{\mathcal{H}_0, f_0, f_{z0}, f_{2z0}, f_{3z0}\}\,, \\
 \text{C} &=& \{\mathcal{H}_0, f_0, f_{z0}, f_{2z0}, f_{3z0}, f_{4z0}\}\,.
\end{eqnarray}

In particular, the reason for studying the fCS in such a hierarchical way is that it is naively expected a broadening of the sampled distributions by adding more parameters. The corresponding numerical effects to the measured quantities lead to strong error propagations; this is due to the higher orders of the Taylor expansion.
We are interested both in quantifying these effects and in fixing constraints on the fCS. Our numerical study is based on a Monte Carlo simulation, in which the chosen observational datasets for our fits can be summarized as follows

\begin{itemize}
 \item The union 2.1 SNeIa compilation of the supernova cosmology project \cite{Suzuki:2011hu}.
This sample is an update dataset of the previous compilations union 2 \cite{Amanullah:2010vv} and union 1 \cite{Kowalski:2008ez}. Union 2.1 includes measurements in the plane $\mu-z$ of 580 supernovae over the redshift range $ 0.015 < z < 1.414$. In the following numerical analyses, we take into account systematic errors in the covariance matrix.

\item Observations of the Hubble factor (OHD) as a function of redshift. We take the compilation of reference \cite{Moresco:2012by} which encompasses 18 measurements between the redshift range $ 0.09 < z < 1.75 $ (see Tab. I of \cite{Moresco:2012by}). The data are extracted from previous works (see for ex.     \cite{Simon:2004tf,Stern:2009ep,Moresco:2012jh}).

\item A gaussian prior on the Hubble constant of $\mathcal{H}_0 = 74.2 \pm 3.6 \, \text{km/s/Mpc}$ \cite{Riess:2009pu}, as measured by the Hubble Space Telescope (HST).
\end{itemize}

To constrain the parameters, we use a Bayesian method in which the best fits of the parameters are those which maximize the likelihood function

\begin{equation}
 \mathcal{L} \propto \exp(-\chi^2/2)\,,
\end{equation}

where $\chi^2$ is the  ({\it pseudo}){\it chi-squared} function \cite{Hobson2010}. Since the different sets of observations are not correlated, the function $\chi^2$ is simply given by the sum

\begin{equation}
 \chi^2 = \chi^2_{\text{Union2.1}} + \chi^2_{\text{HST}} + \chi^2_{\text{OHD}}\,.
\end{equation}

We perform a Markov Chain Monte Carlo analysis by modifying the publicly available code CosmoMC \cite{Lewis:2002ah} (see also \cite{cosmomc_notes}). To obtain the posterior distributions, we assume uniform priors over the intervals given in Tab. I. In Tab. \ref{table:summary}, we show the summary of the constraints.
We report the best fits given by the maximum of the likelihood function of the samples, the quoted errors
show the $68 \%$ confidence level (c.l.). In Fig. 4 we plot the corresponding posterior distributions.
The vertical lines denote the upper and lower limits for the $\Lambda$CDM case, these are obtained by using the best fits parameters reported in Tab. I, compatible with those of \cite{Komatsu:2010fb}. In Figs. 1, 2 and 3, we show all the 2-dimensional marginalized posterior confidence intervals for the three analyzed models.

As it can be noticed from figures 1, 2, 3 and 4, the marginalized posteriors loose Gaussianity when we add further parameters to Model A. We conclude that considering Model C over Model B has the advantage that it gives more information on the cosmographic $f(\mathcal{R})$ parameters without enlarge the dispersions; however, Model C is less suitable for a posterior statistical treatment.

\begin{figure}
\begin{center}
\includegraphics[width=1.6in]{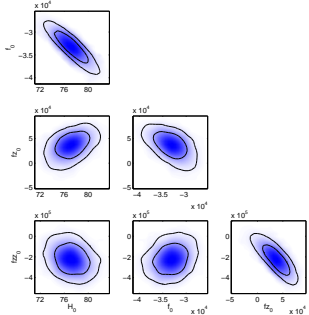}
\caption{2-dimensional marginalized probability for the parameters of model A. The dashing denotes the likelihood of the samples.}
\end{center}
\label{2dpdfA}
\end{figure}

\begin{figure}
\begin{center}
\includegraphics[width=1.6in]{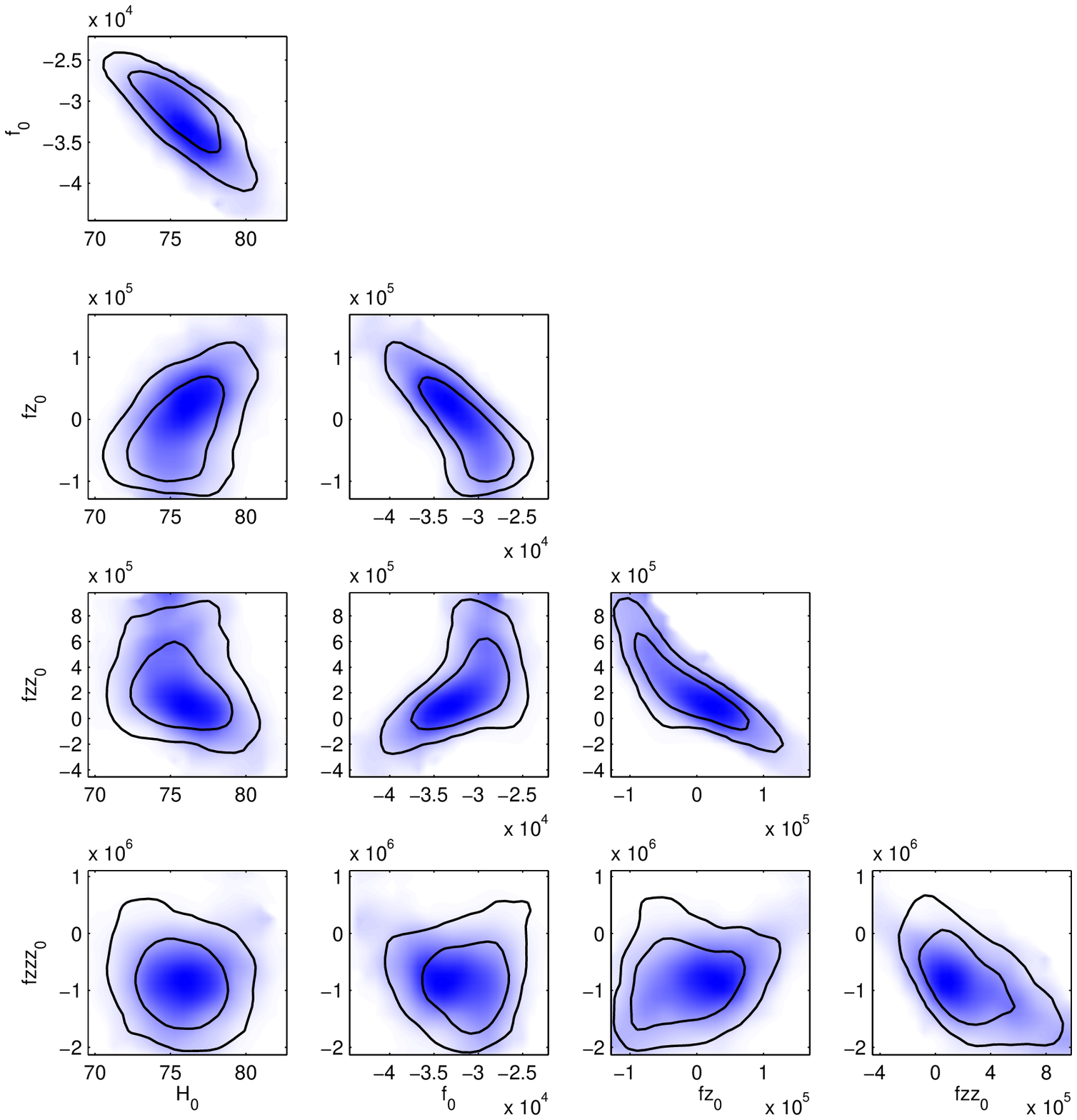}
\caption{2-dimensional marginalized probability for the parameters of model B. The dashing denotes the  likelihood of The samples.}
\end{center}
\label{2dpdfB}
\end{figure}

\begin{figure}[h]
\begin{center}
\includegraphics[width=1.6in]{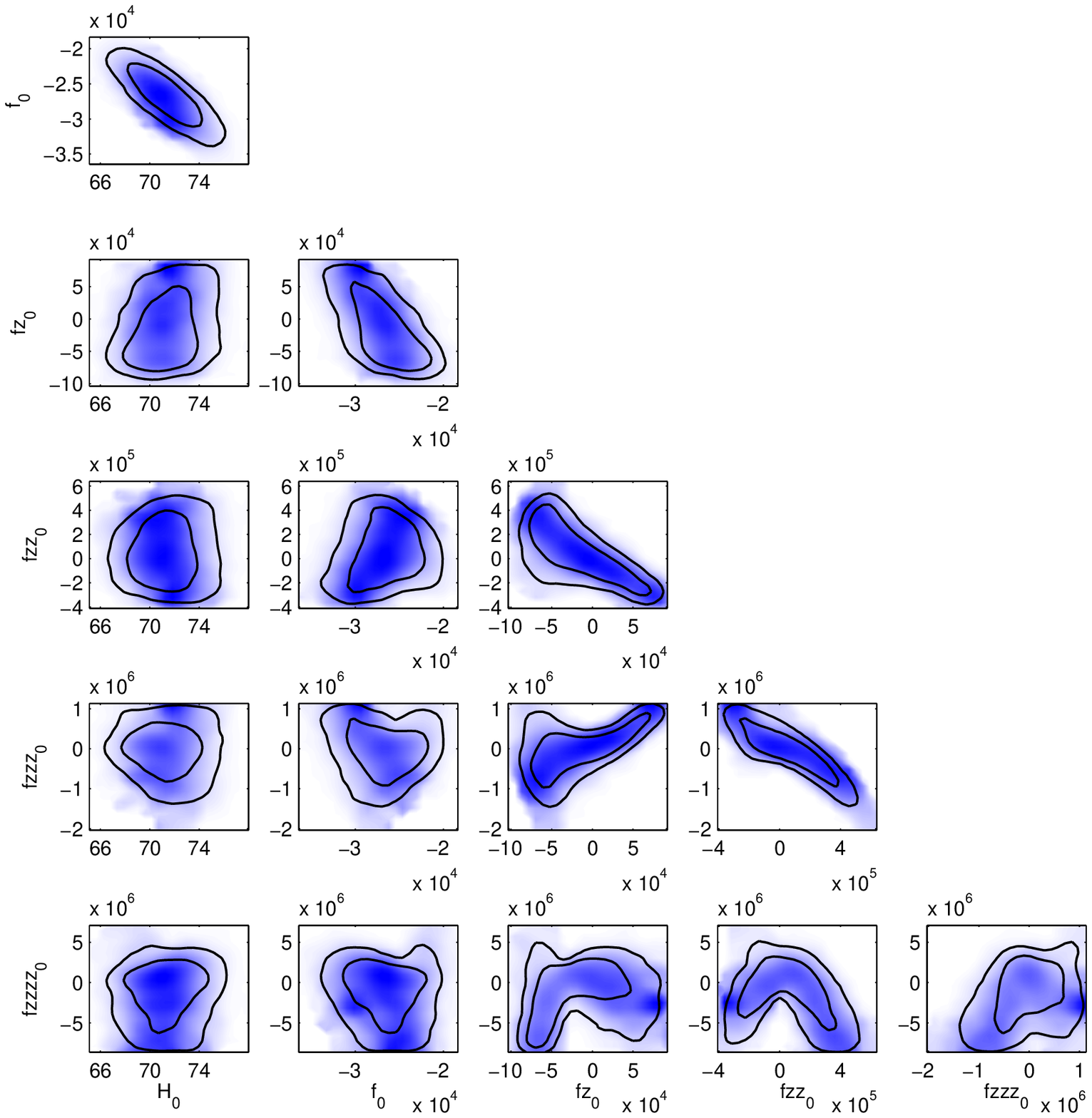}
\caption{2-dimensional marginalized probability for the parameters of model C. The dashing denotes the likelihood of the samples.}
\end{center}
\label{2dpdfC}
\end{figure}

\begin{table*}
\caption{{\footnotesize Best fits of the parameters for the three considered models.  The quoted errors show the $0.68$ c.l.  The observations used
to constrain the parameters are the union2.1 data set compilation, Observational determination of the Hubble factor (OHD), and the measured value of $\mathcal{H}_0$ by the HST.}}
\begin{tabular}{c|c|c|c} 
\hline\hline\hline 
$\qquad${\small Parameter}$\qquad$   &  $\qquad$ Model A $\qquad$& $\qquad$ Model B $\qquad$ & $\qquad$ Model C $\qquad$ \\
                                          & $\chi^2_{min}= 529.0 $  &  $\chi^2_{min}= 540.0 $  & $\chi^2_{min}= 552.6$  \\[1ex]
\hline
$\mathcal{H}_0$                 & {\small $77.23$}{\tiny${}_{-1.82}^{+0.84}$}       & {\small $75.69$}{\tiny${}_{-1.99}^{+2.03}$}       & {\small $71.30$}{\tiny${}_{-1.91}^{+1.92}$}     \\[0.8ex]
\hline
$10^{-4} f_0$         & {\small $-3.324$}{\tiny${}_{-0.230}^{+0.227}$}    & {\small $-3.144$}{\tiny${}_{-0.332}^{+0.320}$}    & {\small $-2.669$}{\tiny${}_{-0.284}^{+0.287}$}  \\[0.8ex]
\hline
$10^{-4} f_{z0}$      & {\small $3.636$}{\tiny${}_{-1.735}^{+1.751}$}     & {\small $-1.510$}{\tiny${}_{-5.656}^{+5.694}$}    & {\small $-1.794$}{\tiny${}_{-4.200}^{+4.834}$}  \\[0.8ex]
\hline
$10^{-5} f_{2z0}$     & {\small $-2.202$}{\tiny${}_{-0.973}^{+0.965}$}    & {\small $2.276$}{\tiny${}_{-2.032}^{+2.339}$}     & {\small $0.499$}{\tiny${}_{-2.049}^{+2.192}$}   \\[0.8ex]
\hline
$10^{-5} f_{3z0}$    & $--$                                              & {\small $-8.264$}{\tiny${}_{-5.256}^{+5.064}$}    & {\small $-0.399$}{\tiny${}_{-4.628}^{+4.424}$}  \\[0.8ex]
\hline
$10^{-6} f_{4z0}$   & $--$                                              & $--$                                              & {\small $-1.027$}{\tiny${}_{-3.132}^{+2.430}$}  \\[0.8ex]

\hline\hline\hline
\end{tabular}

{\footnotesize
Notes.
$\mathcal{H}_0$ is given in Km/s/Mpc.}
\label{table:summary} 
\end{table*}

\begin{figure}
\begin{center}
\includegraphics[width=1.6in]{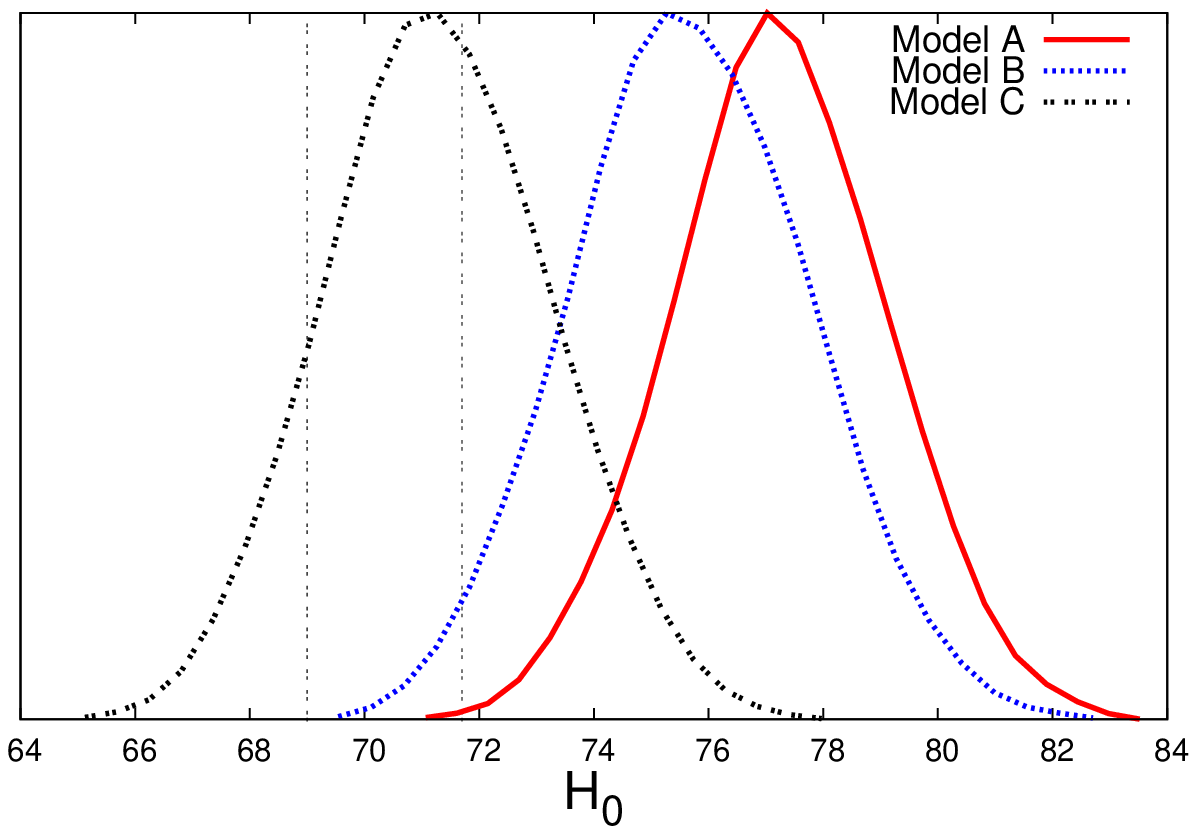}
\includegraphics[width=1.6in]{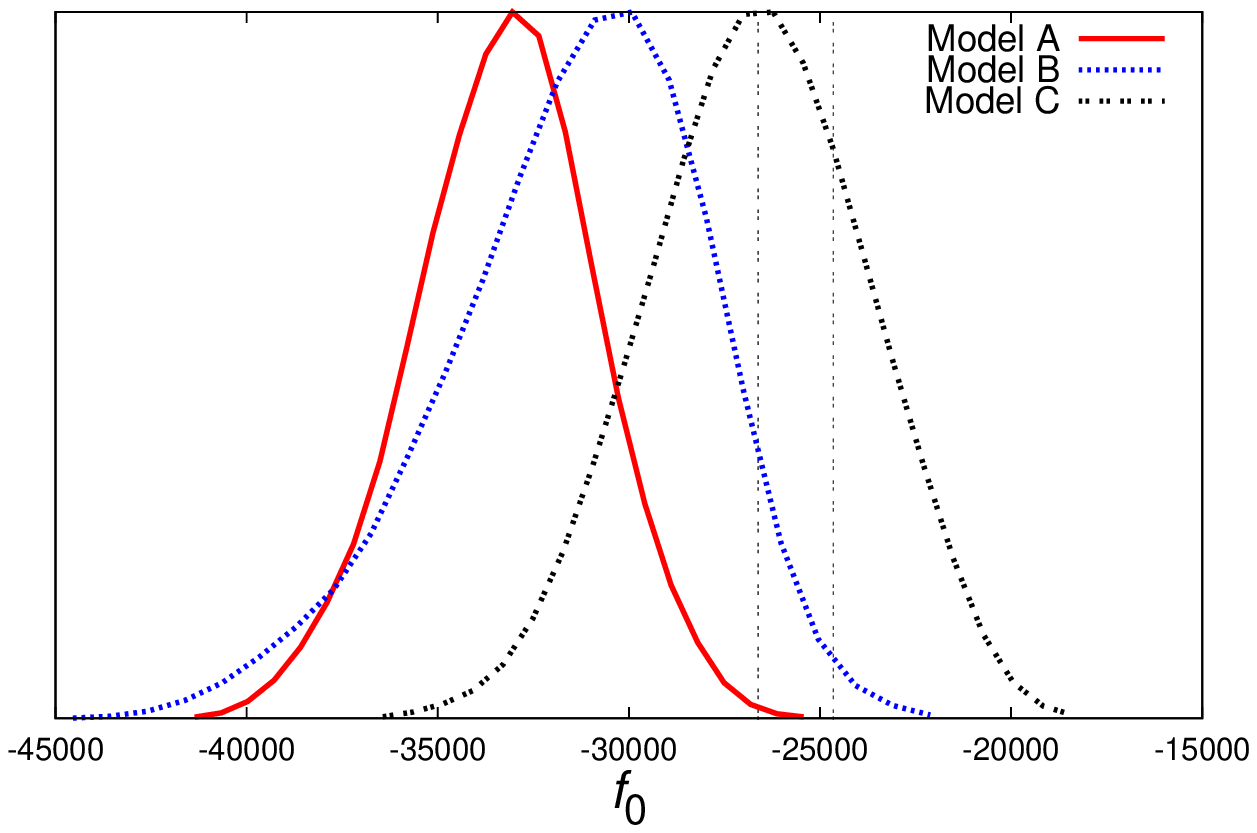}
\includegraphics[width=1.6in]{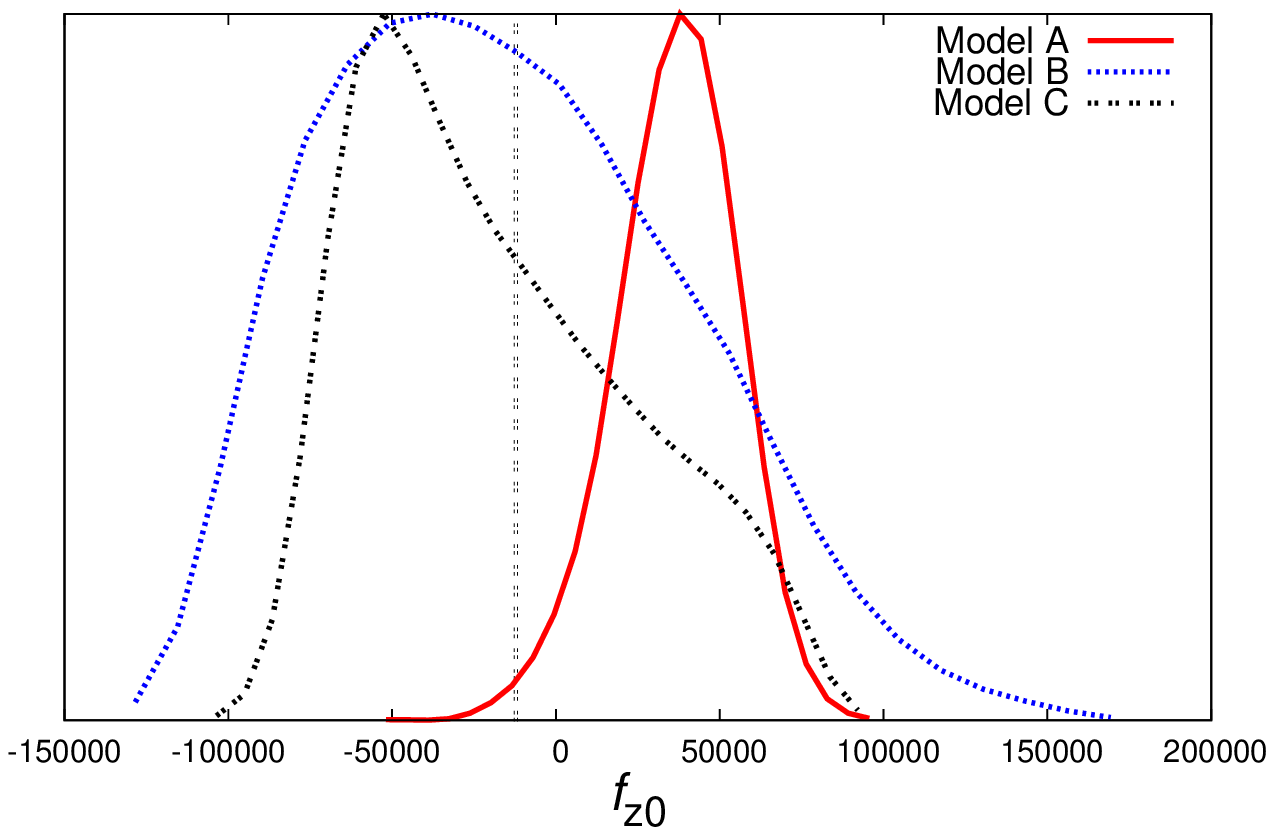}
\includegraphics[width=1.6in]{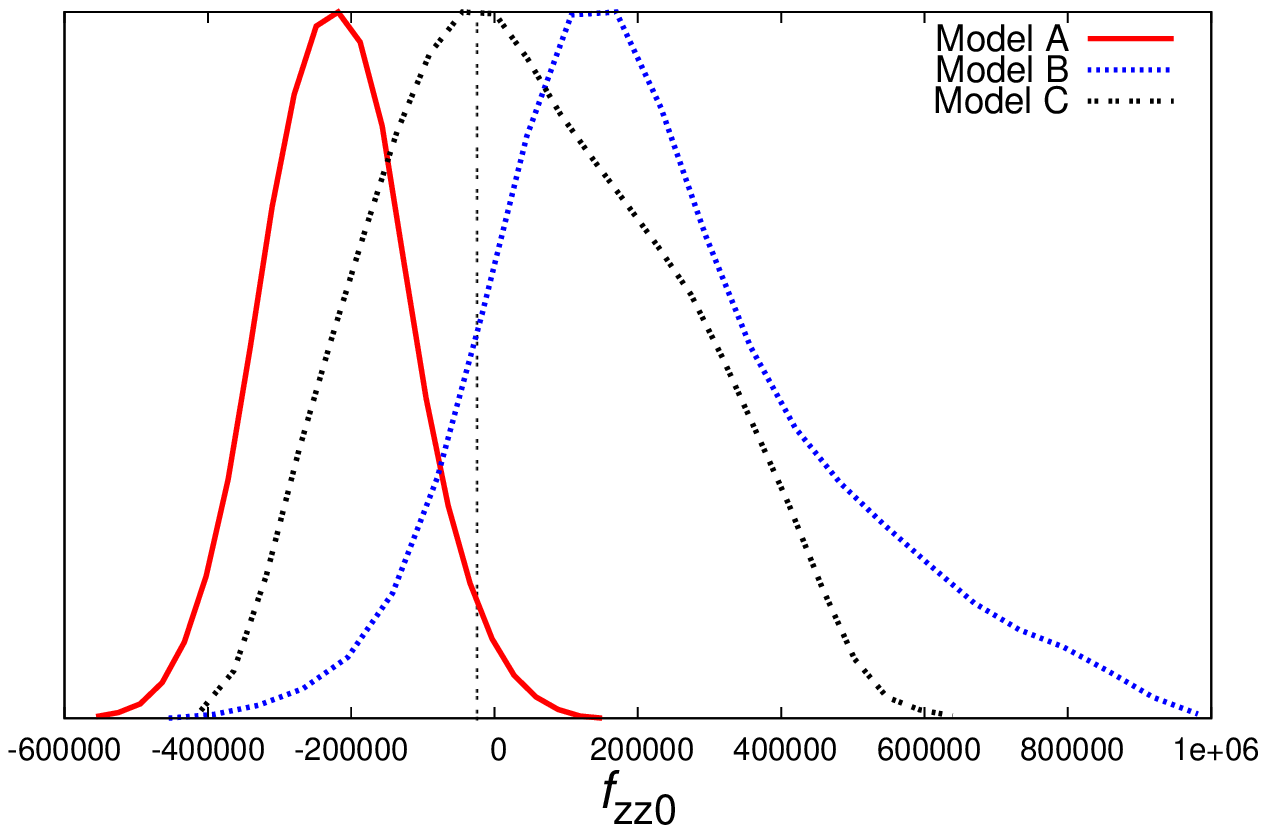}
\includegraphics[width=1.6in]{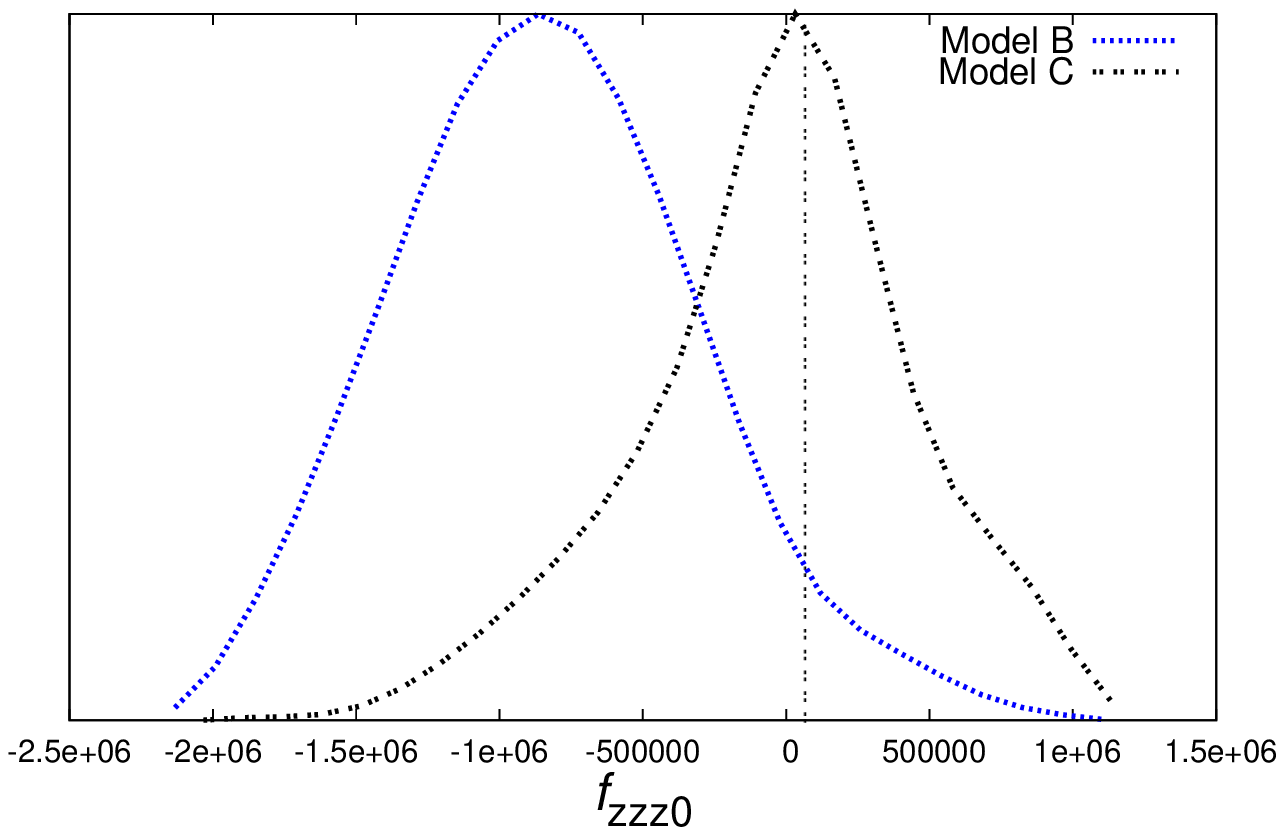}
\includegraphics[width=1.6in]{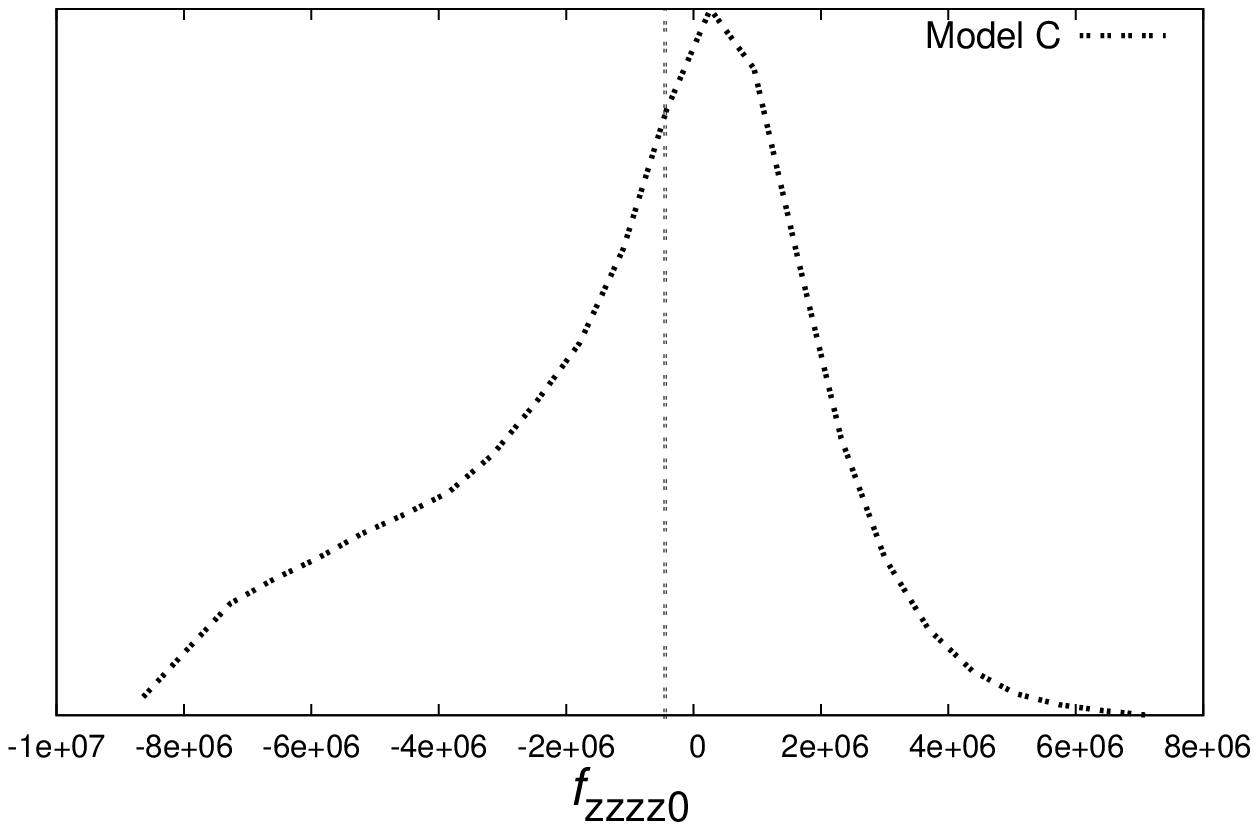}
\caption{1-dimensional marginalized probability for the parameters ex\-plo\-red with MCMC. Solid lines (red) are for model A, dotted lines (blue) for model B,
and dashed (black) for model C. The vertical dashed lines are the lower and upper limits allowed for the $\Lambda$CDM model by using the WMAP7y + BAO + $\mathcal{H}_0$
observations as inferred in \cite{Komatsu:2010fb}. Note that for the cases $f_{2z0}$, $f_{3z0}$ and $f_{4z0}$ these are very close and cannot be distinguised.}
\end{center}
\label{1dpdf}
\end{figure}

\begin{table*}
\caption{{\footnotesize Standard proportions of $f(z)$ derivatives.}}
\begin{tabular}{c|c} 
\hline\hline\hline 
$\qquad${\small Parameter}$\qquad$   &  $\qquad$Standard deviations proportions  $\qquad$\\
                                          &  Model A : Model B : Model C \\[1ex]
\hline
$\mathcal{H}_0$                 & $1:1.51:1.44$         \\[0.8ex]
\hline
$ f_0$         &  $1:1.43:1.25$      \\[0.8ex]
\hline
$ f_{z0}$      &  $1:3.26:2.59$     \\[0.8ex]
\hline
$ f_{2z0}$     &   $1:2.26:2.19$    \\[0.8ex]

\hline\hline\hline
\end{tabular}

\label{table:proportions} 
\end{table*}

\begin{table*}
\caption{{\footnotesize Values of the cosmographic set.}}
\begin{tabular}{c|c|c|c}
\hline\hline\hline
$\qquad${\small Parameter}$\qquad$   &  $\qquad$ Model A $\qquad$& $\qquad$ Model B $\qquad$ & $\qquad$ Model C $\qquad$ \\

\hline
$q_0$         & {$-0.786$}{\tiny${}_{-0.324}^{+0.251}$}    & {$-0.744$}{\tiny${}_{-0.434}^{+0.426}$}    & {$-0.625$}{\tiny${}_{-0.420}^{+0.424}$}  \\[2.3ex]
\hline
$j_0$      & {$2.229$}{\tiny${}_{-0.761}^{+0.718}$}      & {$0.817$}{\tiny${}_{-2.102}^{+2.106}$}    & {$0.787$}{\tiny${}_{-1.83}^{+2.04}$} \\[2.3ex]
\hline
$s_0$     & {$-7.713$}{\tiny${}_{-5.372}^{+4.997}$}      & {$-6.671$}{\tiny${}_{-10.295}^{+11.15}$}    & {$-2.217$}{\tiny${}_{-11.15}^{+11.93}$}  \\[2.3ex]
\hline
$l_0$    & $--$                                              & {$21.003$}{\tiny${}_{-59.593}^{+61.257}$}    & {$9.416$}{\tiny${}_{-58.31}^{+60.72}$} \\[2.3ex]
\hline
$m_0$   & $--$                                              & $--$                                              & {$-41.781$}{\tiny${}_{-432.73}^{+422.23}$}  \\[2.3ex]

\hline\hline\hline
\end{tabular}

{\footnotesize
Table of numerical results for the CS; the numerical values are given at $z=0$, while the error propagations have been found using the standard differential rule.}
\label{table:CS} 
\end{table*}

\begin{table*}
\caption{{\footnotesize Values of $f(R)$ and its derivatives.}}
\begin{tabular}{c|c|c|c} 
\hline\hline\hline 
$\qquad${\small Parameter}$\qquad$   &  $\qquad$ Model A $\qquad$& $\qquad$ Model B $\qquad$ & $\qquad$ Model C $\qquad$ \\

\hline
$f(\mathcal{R}_0)$         & {$-3.324$}{\tiny${}_{-0.230}^{+0.227}$}    & {$-3.144$}{\tiny${}_{-0.332}^{+0.320}$}    & {$-2.669$}{\tiny${}_{-0.284}^{+0.287}$}  \\[2.3ex]
\hline
$f'(\mathcal{R}_0)$      & {$1$}{\tiny${}_{-2.7\cdot10^{-16}}^{+2.6\cdot10^{-16}}$}     & {$1$}{\tiny${}_{-1.8\cdot10^{-15}}^{+1.8\cdot10^{-15}}$}    & {$1$}{\tiny${}_{-5.3\cdot10^{-16}}^{+5.8\cdot10^{-16}}$}  \\[2.3ex]
\hline
$f^{''}(\mathcal{R}_0)$     & {$5.9\cdot10^{-20}$}{\tiny${}_{-7.8\cdot10^{-20}}^{+7.3\cdot10^{-20} }$}     & {$-4.1\cdot10^{-19}$}{\tiny${}_{-4.7\cdot10^{-18}}^{+4.8\cdot10^{-18} }$}     & {$-1.2\cdot10^{-19}$}{\tiny${}_{-6.8\cdot10^{-19}}^{+7.8\cdot10^{-19} }$}   \\[2.3ex]
\hline
$f^{'''}(\mathcal{R}_0)$    & $--$                                              & {$1.8\cdot10^{-9}$}{\tiny${}_{-1.6\cdot10^{-8}}^{+1.6\cdot10^{-8} }$}     & {$1.8\cdot10^{-9}$}{\tiny${}_{-9.5\cdot10^{-9}}^{+1.1\cdot10^{-8} }$} \\[2.3ex]
\hline
$f^{iv}(\mathcal{R}_0)$   & $--$                                              & $--$                                              & {$-3.87\cdot10^{-13}$}{\tiny${}_{-8.2\cdot10^{-12}}^{+9.1\cdot10^{-12} }$}   \\[2.3ex]

\hline\hline\hline
\end{tabular}

{\footnotesize
Table of numerical references for $f(\mathcal{R})$ and its derivatives, evaluated at $z=0$, i.e. $\mathcal{R}=\mathcal{R}_0$; the error propagations have been evaluated through the standard differential rule.}
\label{table:summary2} 
\end{table*}

We note that the dispersions of the samples are considerably enlarged when the third derivative $f_{3z0}$ is included within Model A. In other words, the corresponding Model B suffers from a deep dispersion problem due to the considered dataset of 580 SNeIa. Nonetheless, the introduction of $f_{4z0}$ in Model C does not substantially broaden the distributions. To quantify such an effect, the standard deviations of the distributions are in the proportions given in Tab. \ref{table:proportions}. An additional comment comes from the strong tension between Model A and the $\Lambda$CDM model; such a tension could be substantially alleviated by considering Model B and Model C.

\section{Examples of $f(\mathcal{R})$ gravity}


In this section, we provide a new explicit example of an $f(\mathcal{R})$ model that reduces to $\Lambda$CDM when $z\sim0$ and satisfies the theoretical constraints (\ref{constraint1}) and (\ref{constraint2}). In doing so, we combine recent theoretical results with our cosmographic constraints \cite{nuovo1}. Particularly, several authors recently suggested that viable forms for $f(\mathcal R)$ may be represented by polynomial or exponential functions \cite{nuovo2}. Additional approaches have been proposed in the literature, showing that it is possible to better constrain the cosmological data with further assumptions \cite{nuovo3}. Thus, we set the free parameters of our model according to the new constraints on higher order derivatives that we found in Sec. III from cosmography. In other words, our reconstruction of the $f(\mathcal R)$ function is based on modelling the discrepancies with the data by smoothing different functions, through the use of a Bayesian inverse analysis. The expression for our $f(\mathcal{R})$ candidate is therefore derived in accordance with the above results, through the inverse procedure of determining from data the correct $f(\mathcal{R})$ \cite{eccolo}. Thus, we consider a combination of viable $f(\mathcal R)$ functions, showing that,  in the redshift range $z\preceq 1.41$, our $f(\mathcal R)$ is able to better fit the cosmographic results than previous approaches. We get

\begin{equation}\label{fnostra}
\begin{split}
f(\mathcal{R})=&\frac{1}{2(a+b+c)e\pi \mathcal{R}_0^2}\bigg\{
\Lambda\mathcal{R}_0^2\bigg[\,
2a\pi e^{\mathcal{R}/\mathcal{R}_0}+\\
&+e\bigg(6b+(a+2c)\pi+8b\, \arctan\left(\frac{\mathcal{R}}{\mathcal{R}_0}\right)\bigg)
\bigg]+\\
&+e\mathcal{R}\bigg[
2\mathcal{R}_0\big((a+b+c)\pi \mathcal{R}_0-4b\Lambda\big)+\\
&+(2b-a\pi)\Lambda \mathcal{R}
\bigg]-2ce\pi \Lambda(\mathcal{R}-\mathcal{R}_0)^2\sin\left(\frac{2\pi \mathcal{R} }{\mathcal{R}_0}\right)
\bigg\}\,,
\end{split}
\end{equation}
with $a,b,c$ free parameters of the model.
Clearly, with this choice for $f(\mathcal{R})$ we obtain $f(\mathcal{R}_0)=\mathcal{R}_0+\Lambda$,
$f'(\mathcal{R}_0)=1$ and $f''(\mathcal{R}_0)=0$, independently of the parameters.
Next, we calculate the third and fourth derivatives in $\mathcal{R}=\mathcal{R}_0$, i.e.
\begin{equation}\label{f'''0}
f^{'''}(\mathcal R_0)=\Lambda \,\frac{2b+\pi(a-12c\pi)}{(a+b+c)\pi \mathcal{R}_0^3}\,,
\end{equation}
and
\begin{equation}\label{fiv0}
f^{iv}(\mathcal R_0)=\frac{a\Lambda }{(a+b+c) \mathcal{R}_0^4}\,.
\end{equation}
Again, we use equations (\ref{Rinz0}) and (\ref{Hinz0}) to write $\mathcal{R}_0$ in terms of the CS
and set the value of $\Lambda=2(1-2q_0)\mathcal H_0^2$ (according to $\Lambda$CDM).
Using the numerical values in Tab. \ref{table:CS},
we get the numerical results for
the higher order derivatives of our model, i.e.
 \begin{equation}\label{f'''0numerical}
f^{'''}(\mathcal R_0)=-5.98 \times 10^{-11}\,\frac{2b+\pi(a-12c\pi) }{a+b+c}\,,
\end{equation}
and
\begin{equation}\label{fiv0numerical}
f^{iv}(\mathcal R_0)=\frac{3.79 \times 10^{-15} a }{a+b+c}\,.
\end{equation}
We can compare the results to those in Tab. \ref{table:summary2} (Model C)
to obtain the following constraints on our model
\begin{equation}\label{constrainta}
\begin{split}
a&\sim145.5\,,\\
b&\sim-148\,,\\
c&\sim1\,.
\end{split}
\end{equation}

Equation ($\ref{fnostra}$) represents a first example of $f(R)$, satisfying the cosmographic constraints of fCS. We evaluated Eq. ($\ref{fnostra}$) by using the bounds of Tabs. II and V. We hope that such a choice could represent a viable candidate to extend the $\Lambda$CDM model as a limiting case.

\section{Final forecasts}

In this paper, we addressed the problem of reconstructing the correct form of $f(\mathcal{R})$, through the use of the so called cosmography of the universe. In particular, we considered cosmography as a tool to infer cosmological bounds on $f(z)$ and its derivatives up to the fourth order and consequently on $f(\mathcal{R})$ and its derivatives, at our time. In addition, by considering the class of $f(\mathcal{R})$ which reduces to $\Lambda$CDM at $z\ll1$, we got numerical constraints on $f(\mathcal{R})$ and its derivatives by relating such quantities to the CS.

Once we rewrite the luminosity distance in terms of the $f(\mathcal{R})$ coefficients, we can \emph{directly} measures them, alleviating the problems of error propagation. In particular, we defined such a set of quantities as the fCS, which can be expressed in terms of the well known CS. We found the numerical constraints through the use of Monte Carlo statistical analyses, by adopting the updated union 2.1 dataset, the HST bound for $H_0$ and the OHD measurements.

In this coarse grained picture, we were able to get stringent limits for the fCS and we propose a candidate of $f(\mathcal{R})$, able to reproduce the dynamics of the universe in accordance with the cosmographic results. We hope that the reconstruction of $f(\mathcal{R})$ by using the cosmographic approach can be extended in future works in order to get more relevant constraints on different class of $f(\mathcal{R})$.

\section*{Acknowledgements}

A.A. acknowledges CONACYT for grant no. 215819. A.B. wants to thank prof. H. Quevedo for discussions and ICRA for financial support; O.L. is grateful to dr. C. Gruber for useful comments. This work was supported in part by DGAPA-UNAM, grant No. IN106110.

\begin{widetext}
\appendix

\section{Luminosity distance in terms of the CS and of the fCS}

In this appendix, we write the formulae for the expansion of the luminosity distance $d_L(z)$ in terms of the CS and of the fCS around $z=0$. More details can be found in \cite{mioprdnuovo}. The expansion in terms of the CS of the luminosity distance reads
\begin{eqnarray}\label{sestoordine}
   d_L(z) & = & \frac{1}{\mathcal{H}_0} \Bigl[ z +\frac{1}{2} \Bigl(1-
q_0 \Bigr)  z^2 -\frac{1}{6}
  \Bigl(1-q_0+j_0 -3q_0^2 \Bigr)  z^3  +
 \frac{1}{24} \Bigl( 2 + 5 j_0 -
2q_0 + 10 j_0 q_0 -15 q_0^2(1+q_0) +s_0 \Bigr) z^4  +
\nonumber\\
   & + & \Bigl( -\frac{1}{20} - \frac{9 j_0}{40} +
\frac{j_0^2}{12} - \frac{l_0}{120} +
   \frac{q_0}{20} - \frac{11 j_0 q_0}{12} + \frac{27 q_0^2}{40} - \frac{7
j_0 q_0^2}{8} + \frac{11 q_0^3}{8} +
   \frac{7 q_0^4}{8} - \frac{11 s_0}{120} - \frac{q_0 s_0}{8} \Bigr) z^5  +
\nonumber\\
   & + &
  \Bigl( \frac{1}{30} + \frac{7 j_0}{30} - \frac{19
j_0^2}{72} + \frac{19 l_0}{720} +
   \frac{m_0}{720} - \frac{q_0}{30} + \frac{13 j_0 q_0}{9} - \frac{7
j_0^2 q_0}{18} + \frac{7 l_0 q_0}{240}
   - \frac{7 q_0^2}{10} + \frac{133 j_0 q_0^2}{48} - \frac{13 q_0^3}{6} +
\nonumber\\
   & + & \frac{7 j_0 q_0^3}{4} - \frac{133 q_0^4}{48} - \frac{21
q_0^5}{16} + \frac{13 s_0}{90}
   - \frac{7 j_0 s_0}{144} + \frac{19 q_0 s_0}{48} + \frac{7 q_0^2
s_0}{24} \Bigr)
z^6 +\mathcal{O}(z^7)  \Bigr]\,,
\end{eqnarray}

which is a result evaluated at $k=0$; for extensions see \cite{starobinskievisser}. Inverting the system of Eqs. (\ref{f0fz0fzz0dopo}) to obtain the CS in terms of the fCS, we can rewrite Eq. (\ref{sestoordine}) in terms of the fCS only. We have
\begin{eqnarray}\label{sestoordinebis}
   d_L(z) & = & \frac{1}{\mathcal{H}_0} \Bigl[ z -\frac{f_0+2\mathcal{H}_0^2}{4 \mathcal{H}_0^2} \, z^2+
\frac{9f_0^2+2(36f_0-f_{z0})\mathcal{H}_0^2+108\mathcal{H}_0^4}{72 \mathcal{H}_0^4} \, z^3+
\nonumber\\
   & + &
\frac{-45 f_0^3 + 18 f_0 (-32 f_0 + f_{z0}) \mathcal{H}_0^2 -
 4 (567 f_0 - 21 f_{z0} + f_{2z0}) \mathcal{H}_0^4 - 2592 \mathcal{H}_0^6}{576 \mathcal{H}_0^6}\, z^4+
\nonumber\\
   & + &
\frac{1}{17280 \mathcal{H}_0^8} \Bigl(945 f_0^4 + 2 f_0^2 (8235 f_0 - 274 f_{z0}) \mathcal{H}_0^2 +
   36 (2853 f_0^2 - 141 f_0 f_{z0} + f_{z0}^2 + 4 f_0 f_{2z0}) \mathcal{H}_0^4 +
\nonumber\\
   & + &
   24 (11151 f_0 - 459 f_{z0} + 30 f_{2z0} - f_{3z0}) \mathcal{H}_0^6 + 241056 \mathcal{H}_0^8\Bigr) \, z^5
\nonumber\\
   & + &
\frac{1}{207360 \mathcal{H}_0^{10}} \Bigl(
-8505 f_0^5 + 2 f_0^3 (-93555 f_0 + 3214 f_{z0}) \mathcal{H}_0^2 -
 4 f_0 (398115 f_0^2 - 22252 f_0 f_{z0} + 225 f_{z0}^2+
\nonumber\\
   & + &
 462 f_0 f_{2z0}) \mathcal{H}_0^4 -
 24 \big(271161 f_0^2 + f_{z0} (187 f_{z0} - 10 f_{2z0}) -
    3 f_0 (5480 f_{z0} - 247 f_{2z0} + 5 f_{3z0})\big) \mathcal{H}_0^6 +
\nonumber\\
   & - &
 48 (263844 f_0 - 11478 f_{z0} + 843 f_{2z0} - 39 f_{3z0} + f_{4z0}) \mathcal{H}_0^8 -
 9315648 \mathcal{H}_0^{10}
\Bigr) \, z^6
\Bigr]\,.
\end{eqnarray}

\end{widetext}

\end{document}